\documentclass[aps,prd,reprint,groupedaddress,showpacs]{revtex4-1}

\usepackage{graphicx}

\usepackage{amssymb}
\usepackage{amsmath}
\usepackage{amsfonts}
\usepackage{wasysym}

\begin{document}

\title{Quantum dynamics of the early universe}

\author{V.E. Kuzmichev}
\author{V.V. Kuzmichev}
\affiliation{Bogolyubov Institute for Theoretical Physics,
Nat.~Acad.~of~Sci.~of~Ukraine, Kiev, 03143 Ukraine}

\date{\today}

\begin{abstract}
Quantum gravity may shed light on the prehistory of the universe.
Quantum corrections to gravity affect the dynamics of the expansion of the universe. Their influence is studied
on the example of the exactly solvable quantum model. The corrections to the energy density and pressure lead to the emergence of
an additional attraction (like dark matter) or repulsion (like dark energy) in the quantum system of the gravitating matter and radiation.
The model explains the accelerating expansion (inflation) in the early universe (the domain of comparatively 
small values of quantum numbers) and a later transition from the decelerating expansion to the accelerating expansion
of the universe (the domain of the very large values of quantum numbers) from a single approach.
The generation of primordial fluctuations of the energy density at the expense of the change of sign of
the quantum correction to the pressure is discussed.
\end{abstract}


\maketitle

Quantum gravity may shed light on the prehistory of the universe before the big-bang scenario takes over.
Numerous different models were proposed
to reconcile the classical theory of gravity, based on general relativity, with current astrophysical data \cite{Ben,Hin,Pla}. 
The transition from the classical model of the universe to its quantum analog adds a new element to the theory
in the form of the quantum corrections to the energy density and pressure. The method of constraint system quantization  
can be taken as a basis of quantum theory of gravity suitable for the investigation of cosmological and other quantum gravitational systems. 
In this article, we use quantum geometrodynamics in order to study an influence of quantum corrections
on the dynamics of a quantum gravitational system (universe). For this purpose, it is reasonable to investigate this
problem in terms of exactly solvable cosmological model. 

We will consider an isotropic cosmological model described by the Friedmann-Robertson-Walker metric. 
In the  case of the maximally symmetric geometry, the geometric properties of the system
are determined by a single variable, namely the cosmic scale factor $a$.  The matter sector is taken
in the form of a uniform scalar field $\phi$ (a surrogate of physical fields)
and a perfect fluid which defines the material reference frame enabling one to recognize the instants of time.
The Hamiltonian of the system has the form \cite{Kuz08,Kuz08a,Kuz08b,Kuz13,Kuz15}
\begin{eqnarray}\label{1}
    H & = & \frac{N}{2} \left\{-\,\pi_{a}^{2} - a^{2} + a^{4} [\rho_{\phi} + \rho_{\gamma}]\right\} \nonumber \\ 
& + & c_{1}\left\{\pi_{\Theta} - \frac{1}{2}\,a^{3} \rho_{0} s\right\}
+ c_{2}\left\{\pi_{\tilde{\lambda}} + \frac{1}{2}\,a^{3} \rho_{0} \right\},
\end{eqnarray}
where $\pi_{a},\, \pi_{\Theta},\, \pi_{\tilde{\lambda}}$ are the momenta 
canonically conjugate with the variables $a$, $\Theta$ (the thermasy), $\tilde{\lambda}$ 
(the potential for the specific free energy taken with an inverse sign),
$\rho_{\phi}$ is the energy density of matter represented by a scalar field $\phi$,
$\rho_{\gamma}$ is the energy density of a perfect fluid, which is a function of the density of the rest mass 
$\rho_{0}$ and the specific entropy $s$ \cite{Sch70}. 
The $N$, $c_{1}$, and $c_{2}$ play the role of the Lagrange multipliers.

Here and below we use the Planck system of units. 
The length $l_{P} = \sqrt{2 G \hbar / (3 \pi c^{3})}$ is taken as a unit of length and the $\rho_{P} = 3 c^{4} / (8 \pi G l_{P}^{2})$ 
is used as a unit of energy density and pressure, where $G$ is Newton's gravitational constant. The proper time $\tau$ is
measured in units of length. 

The Hamiltonian (\ref{1}) is a linear combination of constraints and thus weakly vanishes, $H \approx 0$.
The variations of the Hamiltonian with respect to $N$, $c_{1}$, and $c_{2}$ give three constraint equations,
\begin{eqnarray}\label{2}
-\,\pi_{a}^{2} - a^{2} + a^{4} [\rho_{\phi} + \rho_{\gamma}] \approx 0, \nonumber \\
\pi_{\Theta} - \frac{1}{2}\,a^{3} \rho_{0} s \approx 0, 
\quad \pi_{\tilde{\lambda}} + \frac{1}{2}\,a^{3} \rho_{0} \approx 0.
\end{eqnarray}

From the conservation of these constraints in time, it follows that the number of particles of a perfect fluid in the proper 
volume $\frac{1}{2} a^{3}$ (or $2 \pi^{2} a^{3}$, if $a$ is taken in units of length) 
and the specific entropy conserve: $\frac{1}{2} a^{3} \rho_{0} = \mbox{const}$, $s =  \mbox{const}$.
Taking into account these conservation laws and vanishing of the momenta conjugate with the variables $\rho_{0}$ and $s$,
one can discard degrees of freedom corresponding to these variables, and convert the second-class constraints into
first-class constraints \cite{Kuz08,Kuz08a,Kuz08b}, in accordance with Dirac's proposal.
In quantum theory, first-class constraint equations become constraints on the state vector 
and, in this way, define the space of physical states \cite{Dir64}.

It is convenient to choose the perfect fluid with the density $\rho_{\gamma}$ in the form of relativistic matter (radiation).
Then, one can put $a^{4} \rho_{\gamma} \equiv E = \mbox{const}$. Passing from classical variables in 
constraint equations (\ref{2}) to corresponding operators, we obtain the quantum constraint equations, which vanish when applied 
to the state vector $\langle a, \phi | \Psi (T) \rangle$, where $T$ is the ``arc-parameter measure of time'' connected with the proper time 
$\tau$ by the differential equation $d \tau = a dT$ \cite{Kuz13,Kuz15}. The solution of these equations gives the evolution of
$| \Psi (T) \rangle$ in the time parameter $T$ in the form
\begin{equation}\label{3}
| \Psi  (T) \rangle= e^{i \frac{2}{3} E(T - T_{0})} | \Psi  (T_{0}) \rangle,
\end{equation}
where $T_{0}$ is an arbitrary constant taken as a time reference point. The vector $| \Psi  (T_{0}) \rangle\equiv | \psi  \rangle$
satisfies the equation
\begin{equation}\label{4}
\left( - \partial_{a}^{2} + a^{2} - 2 a \hat{H}_{\phi} \right) | \psi \rangle = E | \psi \rangle.
\end{equation}
The operators
\begin{equation}\label{5}
\hat{H}_{\phi} = \frac{1}{2} a^{3} \hat{\rho}_{\phi}, \quad
\hat{L}_{\phi} = \frac{1}{2} a^{3} \hat{p}_{\phi}
\end{equation}
are the operators of Hamiltonian and Lagrangian of the scalar field $\phi$, where
\begin{equation}\label{6}
    \hat{\rho}_{\phi} = - \frac{2}{a^{6}}\,\partial_{\phi}^{2} + V(\phi), \quad \hat{p}_{\phi} = - \frac{2}{a^{6}}\,\partial_{\phi}^{2} - V(\phi),
\end{equation}
are the operators of energy density and pressure respectively, $V(\phi)$ is the potential term. 

In the case $E = 0$, Eq.~(\ref{4}) reduces to the Wheeler-DeWitt equation of a `minisuperspace model' of the universe filled 
with a uniform scalar field.

The Hamiltonian $\hat{H}_{\phi}$ can be diagonalized by introducing the complete set of orthonormalized functions $\langle \chi |u_{k} \rangle$ 
of quantum scalar field in the representation of generalized variable $\chi = \chi (\frac{1}{2} a^{3}, \phi)$. 
The explicit form of $\chi$ is determined by the form of the potential 
$V(\phi)$ \cite{Kuz13}. For example, in the model $V(\phi) = \lambda_{\alpha} \phi^{\alpha}$, where $\lambda_{\alpha}$ is the coupling
constant and $\alpha \geq 0$, we have $\chi = (2 \lambda_{\alpha})^{\frac{1}{2 + \alpha}} \left(\frac{a^{3}}{2} \right)^{\frac{2}{2 + \alpha}} \phi$.

After averaging over quantum states $|u_{k} \rangle$, the scalar field turns into matter characterized by the energy density $\rho_{m}$
and pressure $p_{m}$,
\begin{equation}\label{8}
\rho_{m} = \langle u_{k} | \hat{\rho}_{\phi}| u_{k} \rangle, \quad
p_{m} = \langle u_{k} | \hat{p}_{\phi}| u_{k} \rangle.
\end{equation} 
The expectation value of the Hamiltonian is
\begin{equation}\label{7}
\langle u_{k}| \hat{H}_{\phi} |u_{k'} \rangle = M_{k} (a) \delta_{k k'},
\end{equation}
where the index of the state $k$ can take both discrete and continuous values, $M_{k} (a) = \frac{1}{2} a^{3} \rho_{m}$ is the proper energy of 
matter in the volume $\frac{1}{2} a^{3}$ with the energy density $\rho_{m}$.

Calculating the expectation values (\ref{8}), we obtain
\begin{equation}\label{9}
\rho_{m} = \frac{2 M_{k}(a)}{a^{3}}, \quad
p_{m} = w_{m} \rho_{m},
\end{equation}
where the equation-of-state parameter $w_{m}$ is equal
\begin{equation}\label{10}
w_{m} = - \frac{1}{3} \frac{d \ln M_{k} (a)}{d \ln a}.
\end{equation}

In the general case, the proper energy $M_{k} (a)$ depends on $a$. It describes a classical source (as a mass-energy) 
of the gravitational field.

In the specific case, in the model $V(\phi) = \lambda_{\alpha} \phi^{\alpha}$, matter reduces to the barotropic fluid with the
parameter $w_{m} = \frac{\alpha - 2}{\alpha + 2}$, which does not depend on $a$. For $\alpha = 0$, the barotropic fluid takes the form
of the vacuum of the scalar field in the $k$th state. The value $\alpha = 1$ corresponds to the strings. Matter in the form of dust is reproduced
by $\alpha = 2$, whereas $\alpha = 4$ leads to the relativistic matter and so on. The so-called stiff Zel'dovich matter is obtained in the
limiting case $\alpha = \infty$.

Using Eq.~(\ref{7}), one can integrate Eq.~(\ref{4}) with respect to the matter field variable.
Expressing the vector $| \psi \rangle$ in the form of expansion in terms of the complete set of states $| u_{k} \rangle$,
\begin{equation}\label{11}
|\psi \rangle = \sum_{k} |u_{k} \rangle \langle u_{k}  |\psi \rangle,
\end{equation}
from Eq.~(\ref{4}), we obtain the equation for the function $\langle a | f_{k} \rangle \equiv \langle u_{k}  |\psi \rangle$,
\begin{equation}\label{12}
\left( - \partial_{a}^{2} + a^{2} - 2 a M_{k} (a) \right) | f_{k} \rangle = E | f_{k} \rangle.
\end{equation}
This equation is an eigenvalue equation. Its solution $| f_{k} \rangle$ is an eigenfunction corresponding to the eigenvalue $E$.
The function $| f_{k} \rangle$ describes the geometrical properties of the quantum universe filled with matter, whose mass-energy
is $M_{k}(a)$.

In order to turn to the classical observables, we shall look for the solution of Eq.~(\ref{12}) in the form
\begin{equation}\label{13}
\langle a | f_{k} \rangle = \frac{C_{k}}{\sqrt{\partial_{a} S_{k}(a)}} e^{i S_{k}(a)}.
\end{equation}
where the phase $S_{k}$ can, generally speaking, be complex and $C_{k}$ is the constant determined by the boundary condition on the function 
$\langle a | f_{k} \rangle$, e.g. on the asymptotics $a \rightarrow \infty$. 

The phase $S_{k}(a)$ satisfies the non-linear equation
\begin{equation}\label{15}
(\partial_{a} S_{k})^{2} + a^{2} - 2 a M_{k}(a) - E = \frac{3}{4} \left(\frac{\partial_{a}^{2} S_{k}}{\partial_{a} S_{k}} \right)^{2} 
- \frac{1}{2} \frac{\partial_{a}^{3} S_{k}}{\partial_{a} S_{k}}.
\end{equation}
This equation is exact. It is equivalent to Eq.~(\ref{12}).
In the classical limit, the right-hand side of Eq.~(\ref{15}) vanishes (it is proportional to $l_{P}^{4}$ in physical units) and this
equation turns into the Hamilton-Jacobi equation for the action $S_{k}(a)|_{\hbar = 0} = S_{k}^{\mbox{\tiny class}}(a)$ and the momentum
$\partial_{a} S_{k}^{\mbox{\tiny class}}(a) = - a \dot{a}$, where a dot is used to denote the derivative with respect to the proper time $\tau$. 
This simple equation modifies, when quantum gravity effects are taken into account. 

Using Eqs.~(\ref{11}) and (\ref{13}), from the equation
of motion $\langle \psi | -i \partial_{a} | \psi \rangle = \langle \psi | -\frac{da}{dT} | \psi \rangle$ \cite{Kuz13,Kuz15}, we get the operator equality
\begin{equation}\label{14}
\partial_{a} S_{k} + \frac{i}{2} \frac{\partial_{a}^{2} S_{k}}{\partial_{a} S_{k}} = -\frac{da}{dT}.
\end{equation}
From this equation, it follows that, generally speaking, the scale factor $a$ can be complex. 
The possibility of an introduction of a complex metric tensor and its relation
to real physical gravitational field was studied \cite{Kuz15,Mof,Cha}. Taking the common point of view,
we shall assume that the physical gravitational field is described by the real part of the metric.

Using Eqs.~(\ref{13}) and (\ref{14}), Eq.~(\ref{15}) can be rewritten in the form of the 
Einstein-Friedmann equation
\begin{equation}\label{16}
\left(\frac{\dot{a}}{a} \right)^{2} = \rho_{m} + \rho_{\gamma} + \rho_{q} - \frac{1}{a^{2}}.
\end{equation}
Differentiating Eq.~(\ref{16}) with respect to the proper time $\tau$ and taking into account the first law of thermodynamics, we have
\begin{equation}\label{17}
\frac{\ddot{a}}{a} = -\frac{1}{2} \left[\rho_{m} + \rho_{\gamma} + \rho_{q} + 3\left(p_{m} + p_{\gamma} + p_{q}  \right)\right].
\end{equation}
Here $\rho_{m}$ and $p_{m}$ are the energy density and pressure of matter with the equation-of-state parameter (\ref{10}), $\rho_{\gamma}$ and $p_{\gamma}$ are the energy density and pressure of radiation
with the equation of state $p_{\gamma} = \frac{1}{3} \rho_{\gamma}$, $\rho_{q}$ and $p_{q}$ are the quantum corrections to the total 
energy density and pressure, which have the form
\begin{equation}\label{18}
\rho_{q} = \frac{Q_{k}(a)}{a^{4}} \equiv \frac{2 M_{Q}(a)}{a^{3}}, \quad
p_{q} = w_{q} \rho_{q},
\end{equation}
where $M_{Q}(a) = \frac{1}{2} a^{3} \rho_{q}$ is the proper energy of the quantum source of the gravitational field and the equation-of-state parameter is equal to
\begin{equation}\label{19}
w_{q} = \frac{1}{3}\left(1 - \frac{d \ln Q_{k} (a)}{d \ln a} \right).
\end{equation}
In Eq.~(\ref{19}), the first term is a correction for relativity, while the second one comes from the quantum dynamics of the system and
it is expressed via the function of the gravitational quantum source
\begin{equation}\label{20}
Q_{k}(a) = i \partial_{a}^{2} S_{k} + \frac{1}{2} \left[\left(\frac{\partial_{a}^{2} S_{k}}{\partial_{a} S_{k}} \right)^{2} 
- \frac{\partial_{a}^{3} S_{k}}{\partial_{a} S_{k}}\right].
\end{equation}
The equations (\ref{16}) and (\ref{17}), rewritten in physical units, demonstrate that the energy density $\rho_{q}$ and pressure $p_{q}$
contain the terms proportional to $l_{P}^{2}$ (from the first term on the right-hand side of Eq.~(\ref{20})) and $l_{P}^{4}$ 
(from the term in square brackets).

The function of the gravitational quantum source $Q_{k}$ is a real-valued function, only when the phase 
can be represented in the form $S_{k} = i S_{k}^{\mbox{\tiny Euclid}}$ with the real-valued Euclidean action. 
In other cases, this function is complex-valued. The imaginary part of
the gravitational quantum source can be responsible for creation and annihilation of particles.

The influence of the gravitational quantum source on the dynamics of the expanding universe depends on
the value and the sign of the corresponding energy density and pressure.

If there exists the domain, where the real-valued function $Q_{k} (a) > 0$ and $\ln Q_{k} (a)$ depends on $\ln a$, 
so that $w_{q}$ can be parametrized in the form
$w_{q} = - \frac{1}{3} \delta$, where $\delta$ is an arbitrary positive or negative constant, then the quantum corrections
can imitate, for example, the contribution from the de Sitter vacuum ($\delta = 3$), domain walls ($\delta = 2$), strings ($\delta = 1$),  
dust ($\delta = 0$), radiation ($\delta =  - 1$), or perfect gas  ($\delta = - 2$). In such a model, the quantum source is
$Q_{k}(a) \sim a^{\delta + 1}$.
Identifying the energy density $\rho_{q} > 0$ with the energy density of dark energy,
one finds that the case $\delta = 3$ reproduces the cosmological constant \cite{Pee}, 
the values $1 < \delta < 3$ correspond to the quintessence \cite{Ste,Ost,Tur},
whereas the phantom field \cite{Cal,Cal03} is described by the values $\delta > 3$.

However, it is possible that quantum effects will generate the quantum corrections, for which the function $Q_{k} (a) < 0$ and
the corresponding energy density is negative. This case is not extraordinary. According to quantum field theory, for instance, 
vacuum fluctuations make a negative contribution to the field energy per unit area (the Casimir effect). As it was shown in Ref.
\cite{Kuz09}, the quantum correction $\rho_{q}$ takes a negative value near initial cosmological singularity.

In this connection, it is appropriate to study an exactly solvable model with the potential of scalar field $V(\phi) = \lambda_{2} \phi^{2}$.
Then mass-energy $M_{k}(a) = \sqrt{2 \lambda_{2}} (k + \frac{1}{2}) \equiv M = const$, where $k = 0,1,2, \ldots$ 
is the number of non-interacting identical particles with the mass $\sqrt{2 \lambda_{2}}$ in the state $| u_{k} \rangle$ of scalar field.
The equations (\ref{12}) and (\ref{15}) have analytical
solutions. The eigenvalue $E$ is determined by the condition of quantization $E = 2 n + 1 - M^{2}$, where $n = 0,1,2, \ldots$ is the quantum 
number which numerates the discrete states of the universe in the potential well $\zeta^{2} \equiv (a - M)^{2}$, and
\begin{equation}\label{21}
\partial_{\zeta} S(\zeta) = i \frac{e^{\zeta^{2}} H_{n}^{-2}(\zeta)}{2 \int_{0}^{\zeta} dx\, e^{x^{2}} H_{n}^{-2}(x)},
\end{equation}
where $H_{n}(\zeta)$ is the Hermitian polynomial, and the index $n$ for $S$ is omitted to simplify the notation. 
The function of the source $Q_{k} (a)$ has the form
\begin{equation}\label{22}
Q(\zeta) = - (2 n + 1) + 2 n\, \frac{H_{n-1}(\zeta) H_{n+1}(\zeta)}{H_{n}^{2}(\zeta)}.
\end{equation}
This function depends on quantum $n$th state of the universe. The condition $\rho_{\gamma} = \frac{E}{a^{4}} \geq 0$
imposes a restriction on the mass $M$ of the universe, $M \leq \sqrt{2 n + 1} $, in accordance
with the quantization condition for $E \geq 0$. The observed part of our universe is characterized by the parameters
$M \sim 10^{61}$ ($\sim 10^{80}$ GeV) and $E \sim 10^{118}$ ($\rho_{\gamma} \approx 10^{-10}$ GeV/cm$^{3}$).
From the viewpoint of the model under consideration, it is in the state with $n \sim 10^{122}$ (up to $\sim 10^{-4}$).
This estimate practically coincides with the estimate given in Ref. \cite{Har}. Therefore, 
one can put $E = M^{2}$ (for the early universe) and $E = 0$ (for the universe with $n \gg 1$) for numerical estimations.

From the explicit form of Eq.~(\ref{22}), it follows that the function of the gravitational quantum source $Q_{k}(a)$ is negative and oscillates 
$n$ times in the interval $0 \leq a < \infty$. This function and its first derivative with respect to $a$ diverge at the points, where the Hermitian 
polynomial $H_{n}(\zeta)$ vanishes. In order to get rid of these divergences, we shall use the following regularization procedure,
\begin{equation}\label{23}
\frac{1}{H_{n}^{\alpha}} \rightarrow \frac{1}{H_{n}^{\alpha}} R_{n},
\end{equation}
where an alternating series is introduced,
\begin{equation}\label{24}
R_{n} = \sum_{\nu = 0}^{\infty} (-1)^{\nu} \left(\frac{\varepsilon_{n}}{H_{n}^{\alpha}} \right)^{\nu} = 1 - 
\frac{\varepsilon_{n}}{H_{n}^{\alpha}} + \ldots.
\end{equation}
Here $\varepsilon_{n}$ is some regularization parameter chosen so that the positions and magnitudes of maxima of the function $Q_{k}(a)$
do not change. The series (\ref{24}) is summed
\begin{equation}\label{25}
R_{n} = \left(1 +  \frac{\varepsilon_{n}}{H_{n}^{\alpha}} \right)^{-1}.
\end{equation}

In Fig.~1 for illustration purposes, it is shown the quantum correction $\rho_{q}$ as a function of $a$ with the 
specific values $n = 2$, $M = 1.58$, $E = M^{2}$, which correspond to the early universe, and $\varepsilon_{n} = 2 (2n + 1)$. 
The main features of the behaviour of $\rho_{q}$ remain intact up to the values $n \gg 1$ and $E \ll M^{2}$.
It is multiplied on $a^{4}$ in order to 
exclude the divergence in the point of the initial cosmological singularity $a = 0$ from the consideration. This quantum correction
to the total energy density in Eqs.~(\ref{16}) and (\ref{17}) is found to take only negative values. When the number $n$ increases,
the number of oscillations increases as well.

In the Fig.~1, one can see that $Q_{k}(0) = const$ and $Q_{k}(a) \rightarrow -1$ at $a \rightarrow \infty$. The local maximum corresponds
to the point $a = M$.

\begin{figure}
\includegraphics[width=8cm]{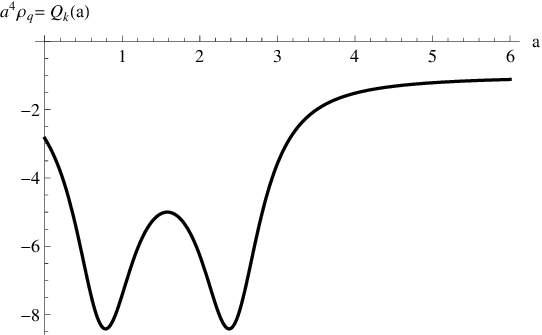}
\caption{\label{fig:1} The quantum corrections $\rho_{q}$ (\ref{18}) to the total energy density multiplied on $a^{4}$ 
for the quantum source $Q_{k}$ (\ref{22}) \textit{versus} the scale factor $a$ for $n = 2$ and $M = 1.58$.}
\end{figure}

The quantum correction $p_{q}$ to the pressure is shown in Fig.~2. This quantum correction $p_{q}$
takes negative values near the origin $a = 0$ and at large values of $a$. It passes through the point $p_{q} = 0$ at $a < 1$ and then takes
positive values. The pressure changes from positive to negative near the point $a = M$, it changes from negative to positive 
near $a = 2M$ and again switches from positive to negative after $a = 4M$. 

\begin{figure}
\includegraphics[width=8cm]{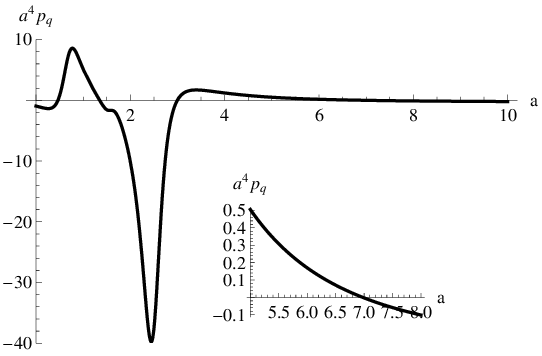}
\caption{\label{fig:2} The quantum corrections $p_{q}$ (\ref{18}) to the pressure multiplied on $a^{4}$ for the quantum source $Q_{k}$ (\ref{22}) 
\textit{versus} the scale factor $a$ for $n = 2$.}
\end{figure}

\begin{figure}
\includegraphics[width=8cm]{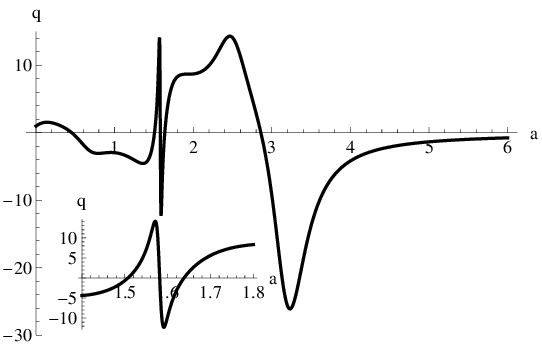}
\caption{\label{fig:3} The deceleration parameter $q$ for the quantum source $Q_{k}$ (\ref{22}) 
\textit{versus} the scale factor $a$ for $n = 2$.}
\end{figure}

\begin{figure}
\includegraphics[width=8cm]{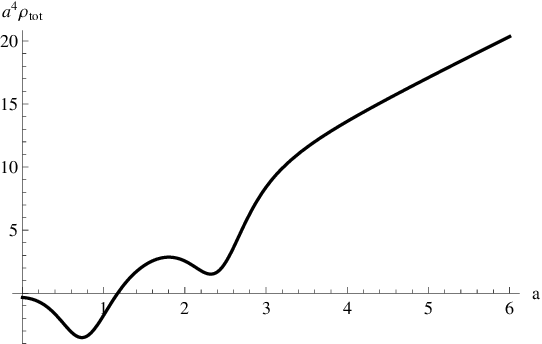}
\caption{\label{fig:4} The total energy density $\rho_{tot}$ multiplied on $a^{4}$ for the quantum source $Q_{k}$ (\ref{22}) 
\textit{versus} the scale factor $a$ for $n = 2$.}
\end{figure}

The peculiarities of the behaviour of the pressure $p_{q}$
manifest themselves in the behaviour of the deceleration parameter $q = - \frac{a \ddot{a}}{\dot{a}^{2}}$ drawn in Fig.~3.
It appears that when one takes into account the quantum corrections for the universe filled with dust and radiation, 
the universe changes from deceleration (near the initial cosmological singularity at
$a = 0$) to accelerating expansion for small $a$. The deceleration parameter takes negative and positive values.
It changes from positive to negative near the points $a = M$ and $a = 2M$. The deceleration parameter
describes the accelerating universe for $a < M$ and for $a > 2 M$ up to the end. The acceleration decreases as the scale factor $a$ increases.
Extrapolating this solution for the case of larger values of
$n$ we come to the conclusion that the quantum corrections generate an additional attraction in the universe acting as dark
matter or a repulsion which can be associated with dark energy. The quantum corrections can, in principle, 
produce both the inflationary expansion in the early universe ($n > 1$) and the accelerating expansion in later times ($n \sim 10^{120}$). 

The way how the quantum corrections exhibit themselves on the total energy density $\rho_{tot} = \rho_{m} + \rho_{\gamma} + \rho_{q}$
can be seen in Fig.~4. The two minima are given by the minima of the quantum correction $\rho_{q}$ shown in Fig.~1. The first minimum
corresponds to the case, when the quantum correction exceeds the contribution from the ordinary matter to the total energy density
in the very early universe. The second minimum demonstrates the predominance of the ordinary matter over the contribution from 
quantum effects in the later era. A bulge of a curve near $a = M$ can be considered as a 
density fluctuation stipulated by corresponding growth of the pressure $p_{q}$ from negative values to positive values for $a < M$
and subsequent change to negative values near $a = M$, shown in Fig.~2. In general case, there are $n - 1$ fluctuations 
for any value of $n$. Such fluctuations may play the role of seeds for future large-scale structure.

In the domain $a > 2 M$, the quantum correction $\rho_{q}$ has the form $\rho_{q} = -\frac{\sigma^{2}}{a^{6}}$,
where $\sigma^{2}$ weakly depends on $a$ (see Fig.~1). 
As it is shown in Ref. \cite{Kuz13}, $\sigma^{2} = \frac{3}{8} N^{2}$, where $N$ is the number
of particles with spin $\frac{1}{2}$ of the Weyssenhoff fluid in the volume $\frac{1}{2} a^{3}$ \cite{Wey,Wey58}. 
The term with the similar dependence on $a$
arises in the Einstein-Cartan theory \cite{Heh,Nur}, where $\rho_{q} = -\langle s_{\mu \nu} s^{\mu \nu} \rangle$, 
the space-time averaging is performed in order to make a transition to macroscopic scales, $s_{\mu \nu}$ is the spin density 
and it is assumed that the average $\langle s_{\mu \nu} \rangle = 0$.

\end{document}